\documentclass[english,twoside,a4paper,10pt]{article}

\usepackage[latin1]{inputenc}
\usepackage[T1]{fontenc}
\usepackage{amsmath}
\usepackage{amsfonts}
\usepackage{graphicx}
\usepackage{subfigure}
\usepackage{a4wide}
\usepackage{amssymb}
\usepackage{fancyhdr}
\usepackage{mathrsfs}
\usepackage[toc,page]{appendix}

\begin{document}

\title{\bf Fluid inflation with brane correction}
\author{ 
Ratbay Myrzakulov\footnote{Email: rmyrzakulov@gmail.com},\,\,\,
Lorenzo Sebastiani\footnote{E-mail address: l.sebastiani@science.unitn.it
}\\
\\
\begin{small}
Eurasian International Center for Theoretical Physics and  Department of General
\end{small}\\
\begin{small} 
Theoretical Physics, Eurasian National University, Astana 010008, Kazakhstan
\end{small}\\
}

\date{}

\maketitle


\begin{abstract}
In this paper, we have investigated the possibility to have inflation from inhomogeneous viscous fluids by taking into account the brane correction coming from string-inspired five dimensional Einsten's gravity. We have realized several kinds of viable solutions for early-time acceleration. At the end of inflation, the classical Einstein's gravity is recovered and fluids produce decelerated expansion.
\end{abstract}



\tableofcontents
\section{Introduction}

The accelerated expansion of the universe today~\cite{WMAP}, and other evidences related to the inflation, namely the early-time acceleration that the universe underwent at the time of the Big Bang~\cite{Linde, revinflazione},
brought physicists to propose a large number of models and scenarios to modify the dynamics of the standard Friedmann universe. Accelerated cosmology may be the result of some  modification of Einstein's gravity~\cite{reviewmod,reviewmod2, myrev}, whose simple form is given by the introduction 
of a positive and small cosmological constant in the framework of General Relativity,  or can be the effect of scalar fields with potential or non-perfect fluids different to standard matter and radiation.

In this paper, we would like to focus our attention on the inflation.
The cosmological observations of the inhomogeneities at the present age give us several informations  about the primordial acceleration, but the choice of the models remains quite large.
The most popular inflation models are realized with a canonical field, the inflaton, subjected to a large potential which induces acceleration~\cite{Guth, Sato}. In the chaotic inflation~\cite{chaotic}, 
the magnitude of the inflaton is initially very large and slowly decreases when the field falls in a potential hole and oscillates, starting the rehating process with the particle production~\cite{buca1, buca2, buca3, buca4}.
Other models are based on higher corrections to Eintein's gravity: in this respect, we would like to mention 
the Starobinsky model~\cite{Staro},
with the account of $R^2$-term in the Hilbert-Eintein action, and whose predictions are in agreement with the Planck data~\cite{Planckdata}: however, the very recent experiments on microwave radiation~\cite{BICEP2} seem to indicate some discrepances with this model.

Actually our proposal, as a prosecution of Ref.~\cite{mioultimo}, is to investigate inflation with inhomogeneous viscous fluids by taking into account the effects of the braneworld inflation coming from higher dimensions string-inspired theories. We will restrict the analysis to the case where the extra dimension is equal to one, namely we will consider the situation where Einstein's gravity holds true in five dimension leading some corrections in four dimension.
This kind of theories are of great interest and have been investigated in many works~\cite{brane1, brane1bis, brane1tris, brane2, brane2bis, brane3, brane3bis, brane3tris, brane4, brane4bis, braneOd1, braneOd2}.
In this models the fundamental Planck scale in five dimension can be considerably smaller than the Planck (mass) scale in four dimension, $M_{Pl} = 1.2 \times 10^{19} \text{GeV}$, with profound consequences at high energies epoch during the inflation~\cite{Maartens, Maeda, Sami, Sami2}.

The paper is organized as follows. In Section {\bf 2}, we will review the formalism of braneworld inflation, introducing the (first) Friedmann equation in four dimension with the brane correction. An extra term respect to Einstein's gravity appears, such that at high energy the Hubble parameter is proportional to the energy density contents of the universe. A brief summary of the canonical scalar field inflation with brane correction is also presented. Section {\bf 3} is devoted to the study of fluid cosmology with brane correction. Several inhomogeneous viscous fluid models are investigated, in the attempt to analyze their feautures in braneworld inflation. In the specific, exponential inflation, quasi de Sitter inflation and quintessence inflation will be examined. For every case, the computation of the characteristic parameters of inflation will be carried out, and the predictions of the various models will be confronted with the last data coming from cosmological observations.
In section {\bf 4}, we will give some comments about the possibility to unify the braneworld fluid inflation with the late-time acceleration of current universe.
Conclusions and final remarks are presented in Section {\bf 5}.


\section{Braneworld inflation}

Let us briefly review some facts about the brane correction of four dimensional Einstein's gravity in Friedmann-Robertson-Walker background. 
If Einstein's gravity holds true in a five-dimensional hyperspace, with a cosmological constant as a source, and the matter fields are confined to the 3-brane, the four-dimensional Einstein's field equations induced on the brane can be written as~\cite{Maeda}
\begin{equation}
G_{\mu\nu}+\Lambda_\text{eff} g_{\mu\nu}=\kappa^2 T_{\mu\nu}+\left(\frac{8\pi}{M_5^3}\right)^2
\pi_{\mu\nu}-E_{\mu\nu}\,.
\end{equation}
Here, $G_{\mu\nu}=R-g_{\mu\nu} R/2$ is the usual Einstein's tensor in four dimension, $R$ being  the Ricci scalar and $g_{\mu\nu}$ the metric tensor, $\Lambda_\text{eff}$ is the effective cosmological constant and $T_{\mu\nu}$ the stress energy tensor of the brane, $\pi_{\mu\nu}$ is a tensor quadratic in $T_{\mu\nu}$
and $E_{\mu\nu}$ is a projection of the five-dimensional Weyl tensor, which describes the effects of the five dimensional hyperspace graviton freedom degrees on the brane dynamics. In the above expression, the Plank Mass $M_{Pl}$ is encoded in $\kappa^2=8\pi/M_{Pl}^2$, and the five-dimensional Plank scale $M_5$ is related to the Plank Mass as
\begin{equation}
M_{Pl}=\sqrt{\frac{3}{4\pi}}\left(\frac{M_5^2}{\sqrt{\lambda}}\right)M_5\,,\label{p5}
\end{equation} 
where $\lambda>0$ is the 3-brane tension. 
The effective cosmological constant $\Lambda_\text{eff}$ is given  by the cosmological constant $\Lambda_5$ in five dimension and by the 3-brane tension $\lambda$,
\begin{equation}
\Lambda_\text{eff}=\sqrt{\frac{4\pi}{M_5^3}}\left(\Lambda_5+\frac{4\pi}{3M_5^3}\lambda^2\right)\,,
\end{equation}
but following the other proposals on the topic we will set $\Lambda_5
=-4\pi\lambda^2/(3M_5^3)$ such that $\Lambda_\text{eff}=0$, avoiding to discuss the fine-tuning problem.

Let us consider the Friedmann-Robertson-Walker (FRW) metric,
\begin{equation}
ds^2=-dt^2+a(t)^2d{\bf x}^2\,,\label{metric}
\end{equation}
where $a(t)$ is the scale factor of the universe and depends on the cosmological time. The first Friedmann equation is derived as~\cite{Friedmod}
\begin{equation}
H^2=\frac{\kappa^2}{3}\rho\left(1+\frac{\rho}{2\lambda}\right)+\frac{\mathcal E}{a^4}\,,
\label{F}
\end{equation}
where $H=\dot a/ a$ is the Hubble parameter, $\rho$ the energy density content of the universe and $\mathcal E$ an integration constant related to $E_{\mu\nu}$, which brings the five dimensional hyperspace gravitation effects on the brane. Since it looks like a radiation term, we may assume that it is quickly shifted away during inflation and we can put $\mathcal E=0$. The second term proportional to $\rho^2/\lambda$ that modifies the classical Friedmann equation derives from the Plank scale in five dimension (\ref{p5}) as $\kappa^2/\lambda=(2/3)(4\pi/(M_5^3))^2$. Such a modification becomes important at inflation scale, when $\lambda\ll\rho$.
In the limit $\lambda\rightarrow\infty$, Equation~(\ref{F}) turns out to be the one of Einstein's theory, and in general we must require $\lambda> (1 \text{MeV})^4$, namely $M_5>10\text{TeV}$, to neglet such a term at the time of nucleosynthesis~\cite{brane4}.

Braneworld inflation has been well studied in literature in the canonical scalar field representation, with a scalar field $\sigma(t)$ depending on the cosmological time and subjected to the potential $V(\sigma)$ and whose energy density and pressure are given by
\begin{equation}
\rho_\sigma=\frac{\dot\sigma^2}{2}+V(\sigma)\,,\quad 
p_\sigma=\frac{\dot\sigma^2}{2}-V(\sigma)\,.
\end{equation}
Here, the dot denotes the derivative with respect to the time.
Since on the brane it must be still valid $\nabla_\mu T^{\mu\nu}$, where $\nabla_\mu$ is the covariant derivative respect to the metric $g_{\mu\nu}$, the field satisfies the energy conservation law,
\begin{equation}
\dot\rho_\sigma+3H(\rho_\sigma+p_\sigma)=0\,,
\end{equation}
namely,
\begin{equation}
\ddot\sigma+3H\dot\sigma=-V'(\sigma)\,,\label{cons}
\end{equation}
being the prime the derivative of the potential with respect to 
the field.
As usually, the acceleration is evaluated as
\begin{equation}
\frac{\ddot{a}}{a}= H^2+\dot{H}=H^2\left(1-\epsilon \right)\,,
\end{equation}
where $\epsilon$ is the so called ``slow roll'' parameter
\begin{equation}
\epsilon=-\frac{\dot{H}}{H^2}\,.\label{epsilon}
\end{equation}
Thus, in order to have an acceleration, one must find $\epsilon<1$. An other important slow roll parameter for inflation is given by
\begin{equation}
\eta=-\frac{\ddot{H}}{2H\dot{H}}=\epsilon-\frac{1}{2 \epsilon H}\dot\epsilon\,.\label{eta}
\end{equation}
The inflation is described by a quasi de Sitter expansion, when the magnitude of the slow roll parameters is very small and the kinetic energy of the field is negligible with respect to the potential. Therefore, the slow roll regime is described by
\begin{equation}
\dot\sigma\ll V(\sigma)\,,\quad \ddot\sigma\ll H\dot\sigma\,.
\end{equation}
 In such a case, the slow roll parameters derived from Equation (\ref{F}) with $\mathcal E=0$ and from Equation (\ref{cons}) in the slow roll limit read
\begin{equation}
\epsilon=\frac{1}{2\kappa^2}\left(\frac{V'(\sigma)}{V(\sigma)}\right)^2
\frac{1+V/\lambda}{(1+V/(2\lambda))^2}\,,
\quad
\eta=\frac{1}{\kappa^2}\left(\frac{V''(\sigma)}{V(\sigma)}\right)
\frac{1}{1+V/(2\lambda)}\,,
\end{equation}
where $\dot H<0$ to permit a graceful exit from inflation at small curvature (i.e. at small Hubble parameter). 
We note that in the absence of the brane world correction it is generally required that $(V'(\sigma)/V(\sigma))^2\,,|V''(\sigma)/V(\sigma)|\ll 1$, but, due to this correction, in the high energy limit when $\lambda\ll V(\sigma)$, the slow roll parameters could be very small even if this terms are large.

At the end of inflation, the kinetic energy of the field must slowly increase according with (\ref{cons}) and the universe exits from the de Sitter expansion when the slow roll parameter $\epsilon$ is on the order of the unit. 

To measure the amount of the de Sitter inflation, we may introduce the number of $e$-folds $N$ as
\begin{equation}
N\equiv\ln \left(\frac{a_\mathrm{f}}{a_\mathrm{i}}\right)
=\int^{t_f}_{t_i} H(t)dt\,,\label{N}
\end{equation}
and for canonical scalar field with brane correction,
\begin{equation}
N\simeq 
\kappa^2\int^{\sigma_i}_{\sigma_e}\frac{V(\sigma)}{V'(\sigma)}\frac{V}{2\lambda}d\sigma
= 
\kappa^2\int^{\sigma_i}_{\sigma_e}\frac{1}{\sqrt{\epsilon}}\sqrt{\frac{V}{2\lambda}}d\sigma\,,
\label{Nscalar}
\end{equation}
where the term $\sqrt{V/(2\lambda)}$ modifies the classical result of scalar field theories in the absence of brane correction.
The inflation solves the problems of initial conditions of the universe (horizon and velocities problems) if $N>76$. 

The amplitude of the primordial scalar power spectrum during inflation is given by
\begin{equation}
\Delta_{\mathcal R}^2=\frac{\kappa^2 H^2}{8\pi^2\epsilon}\,.\label{spectrum}
\end{equation}
Moreover, two important quantities that can be detected today by observing the perturbations in our homogeneous universe are the spectral index $n_s$ and the tensor-to-scalar 
ratio $r$, derived from the slow roll parameters as 
\begin{equation}
n_s=1-6\epsilon+2\eta\,,\quad r=16\epsilon\,.\label{index}
\end{equation}
The Planck satellite results constrain these quatities as 
$n_{\mathrm{s}} = 0.9603 \pm 0.0073\, (68\%\,\mathrm{CL})$ and 
$r < 0.11\, (95\%\,\mathrm{CL})$, but the very recently 
BICEP2 experiment~\cite{BICEP2} has detected the $B$-mode polarization of the cosmic microwave background (CMB) radiation with the tensor to scalar ratio 
as
$r =0.20_{-0.05}^{+0.07}\, (68\%\,\mathrm{CL})$,
and the case that $r$ vanishes 
may be rejected at $7.0 \sigma$ level.

\section{Fluid models for inflation with brane correction}

The general form of the Equation of State (EoS) for an inhomogeneous viscous fluid reads~\cite{fluidOd, fluidOd2, fluidOd3, fluidOd4},
\begin{equation}
p=\omega(\rho)\rho-3 H\zeta(a,H,\dot H, \ddot H,...)\,,\label{EoS}
\end{equation}
where the EoS parameter $\omega(\rho)$ may depend on the fluid energy density (inhomogeneous fluid), and
$\zeta(a, H,...)$ is the viscosity which in general may be a function of the scale factor, the Hubble parameter or of the derivatives of the Hubble parameter (for simplicity, we avoid the dependence of the viscosity on the fluid energy density).
To obtain the positive sign of the entropy change in an irreversible process, $\zeta(a,H,..)$ must be positive~\cite{Alessia, Brevik,Alessia(2)}. 
The energy conservation law ($\nabla_\mu T^{\mu\nu}=0$) for such a kind of fluid is derived as
\begin{equation}
\dot\rho+3H\rho(1+\omega(\rho))=(3H)^2\zeta(a, H, \dot H, \ddot H,...)\,.\label{cons2}
\end{equation}
We are interested to see how inhomogeneous viscous fluids can induce a viable inflation in the presence of the brane correction, namely when Equation~(\ref{F}) with $\mathcal E=0$ holds true. We will assume that during inflation $\lambda \ll\rho$ and 
\begin{equation}
H^2\simeq\frac{\kappa^2\rho^2}{6\lambda}\,,\quad \lambda\ll\rho\,.
\label{F2}
\end{equation} 
On the other side, after the exit from inflation, $\rho\ll\lambda$ and we can find the standard Friedmann cosmology with
\begin{equation}
H^2\simeq\frac{\kappa^2}{3}\rho\,,\quad\rho\ll\lambda\,.\label{Fnormal}
\end{equation}
Thus, to recover the Friedmann universe, we need some reheating for the particle production. The particle production comes from the coupling between the curvature ($R=12H^2+\dot H$) and the quantum fields of the particles and may be induced by damped ocillations of the curvature: in order to obtain such oscillations after the inflation, the conservation law of the fluid must be reduced to an armonic equation, and 
in Ref.~\cite{mioultimo} it has been shown how such a mechanism is possible thanks to the viscosity.
In this paper, we will 
avoid to discuss the rehating after inflation and we will limit to show how universe enters in a decelerated phase.

\subsection{Inflation with exponential Hubble parameter\label{expH}}

Let us consider the following exponential form for the Hubble parameter,
\begin{equation}
H(t)=H_0\text{e}^{-\gamma t}\,,\quad \gamma>0\,,\label{H1}
\end{equation}
where $H_0>0$ is the value of the Hubble parameter when $t=0$ and $\gamma$ is a dimensional constant ($[\gamma]=[H]$). The corresponding scale factor reads
\begin{equation}
a(t)\equiv \exp\left[\int^t H(t) dt'\right]=a_\infty\exp\left[-\frac{H_0\text{e}^{-\gamma t}}{\gamma}\right]\,,\label{a1}
\end{equation}
where $a_\infty>0$ is the value of the scale factor at $t\rightarrow\infty$.
Therefore, the asymptotic limits of $H$ result to be
\begin{equation}
H(t=0)=H_0\,,\quad H(t\rightarrow\infty)=0^+\,.
\end{equation}
This solution is free of initial singularity and inflation starts with the cosmic acceleration at $t_\text{i}=0$.
When $\lambda\ll\rho$, Equation~(\ref{F}) with $\mathcal E=0$ reads as (\ref{F2})
and the universe energy density is proportional to $H$,
\begin{equation}
\rho\simeq\sqrt{\frac{6\lambda}{\kappa^2}}H_0\text{e}^{-\gamma t}\,.\label{fluid1}
\end{equation}
A simple way to obtain this result is given by the inhomogeneous EoS for the fluid,
\begin{equation}
p=-\rho+A_0\,, 
\end{equation}
where $A_0$ is a suitable dimensional constant. This choice corresponds to (\ref{EoS}) with $\omega(\rho)=-1+A_0/\rho$ and $\zeta(H)=0$ (inhomogeneous non viscous fluid). Thus, the conservation law (\ref{cons2}) with (\ref{fluid1}) hold true if
\begin{equation}
\rho=-3A_0\log [a/a_\infty]+\rho_0\,,\quad A_0=\frac{\gamma}{3}\sqrt{\frac{6\lambda}{\kappa^2}}\,,\quad\rho_0=0\,,\label{rho1}
\end{equation}
such that $A_0$ is positive. Note that also the energy density of the fluid is positive and the fluid is real, since $a/a_\infty<1$.

The slow roll parameters (\ref{epsilon})--(\ref{eta}) read
\begin{equation}
\epsilon=\frac{\gamma\text{e}^{\gamma t}}{H_0}\,,
\quad\eta=\frac{\gamma\text{e}^{\gamma t}}{2H_0}\,,
\end{equation}
and if
\begin{equation}
\gamma\ll H_0\,,\label{rel1}
\end{equation}
they are very small when $t$ is close to zero ($t\ll 1/\gamma$)\,, 
\begin{equation}
\epsilon\simeq\frac{\gamma}{H_0}\,,
\quad\eta\simeq\frac{\gamma}{2H_0}\,.
\end{equation}
As a consequence, the universe expands in an accelerated way. 

The $\epsilon$ slow roll parameter goes to one at
\begin{equation}
t_\text{f}=\frac{\log [H_0/\gamma]}{\gamma}\,.
\end{equation}
This quantity is positive defined and $0=t_\text{i}\ll t_\text{f}$ when relation (\ref{rel1}) is satisfied, and indicates the duration of the inflation: after this time the universe starts to decelerate. It is easy to understand it by looking for Eq.~(\ref{a1}), from which we obtain
\begin{equation}
\ddot a=a_0\exp\left[-\frac{H_0}{\gamma}\text{e}^{-\gamma t}\right] H_0\text{e}^{-\gamma t}
\left[-\gamma+H_0\text{e}^{-\gamma t}
\right]\,.
\end{equation}
When $t$ is close to zero, we have an acceleration since $\gamma\ll H_0$, but for large value of $t$ the first term on the right side of the equality becomes dominant and the solution is for decelerated expansion.

The $e$-folds number (\ref{N}) reads
\begin{equation}
N\equiv\int^{t_f}_{t_i} H(t)dt=\frac{H_0}{\gamma}\left[\text{e}^{-\gamma t_\text{i}}-\text{e}^{-\gamma t_\text{f}}\right]\simeq \frac{H_0}{\gamma}\,,
\end{equation}
where we have considered $t_\text{i}=0$ and we used the fact that $0\ll t_\text{f}$. As a result, to obtain $76<N$ we must have $76\gamma<H_0$, well satisfying also condition (\ref{rel1}). Note that during inflation
\begin{equation}
\epsilon\simeq\frac{1}{N}\,,\quad \eta\simeq\frac{1}{2N}\,.
\end{equation}
The amplitude of power spectrum (\ref{spectrum}) is given by
\begin{equation}
\Delta^2_\mathcal R=\frac{\kappa^2 H_0^3}{8\pi^2\gamma\text{e}^{3\gamma t}}\,.
\end{equation}
Finally, the spectral indexes (\ref{index}) are derived as
\begin{equation}
n_s=1-\frac{5\gamma\text{e}^{\gamma t}}{H_0}\simeq\frac{5}{N}\,,\quad r=\frac{16\gamma\text{e}^{\gamma t}}{H_0}\simeq\frac{16}{N}\,.
\end{equation}
These quantities satisfy the Plank data results, but if the $e$-folds is not too large, the tensor-to-scalar ratio $r$ may also be in agreement with the last results of the BICEP2 experiment. For example, if $H_0=76\gamma$ with $N=76$, one has $r\simeq 0.211$ at $t=0$.
Such a possibility is a consequence of the fact that $\epsilon\sim 1/N$ instead $\epsilon\sim 1/N^2$ as usually happens in scalar field theories (see Eq.~(\ref{Nscalar})): this result has been found also in Ref.~\cite{mioultimo}, where inflation in standard Einstein's gravity scenario (without brane correction) has been considered. 

After the end of inflation, if $\rho\ll\lambda$ Equation (\ref{F}) with $\mathcal E=0$ turns out to be the usual Friedmann one (\ref{Fnormal}).
The solution of this equation with the energy conservation law of the fluid under consideration is
\begin{equation}
H=H_1-\frac{A_0 t\kappa^2}{2}\,,
\end{equation}
where $H_1$ is a constant and in general $H_1>A_0\kappa^2 t/2$ for expanding universe. 
Moreover, it is easy to see that as soon as $2(H_1-A_0 t\kappa^2/2)^2<A_0\kappa^2$ the solution remains for decelerating universe. In order to recover the Friedmann cosmology, in this phase the rehating processes must take place or must had taken place before.

\subsection{Quasi de Sitter inflation\label{QuasidS}}

In this subsection, we will consider the following form of the Hubble parameter~\cite{davood},
\begin{equation}
H(t)=H_0+\gamma(t_\text{i}-t)-\delta(t_\text{i}-t)^2\,,\quad\gamma\,,\delta>0\,,\label{H2}
\end{equation}
which is a perturbation with respect to the de Sitter solution $H=H_0$. Here, $H_0>0$ is the value of the Hubble parameter when $t=t_\text{i}$, $t_\text{i}>0$ being the fixed time at the beginning of inflation, and
$\gamma\,,\delta$ are positive dimensional constants ($[\gamma]=[H^2]\,,[\delta]=[H^3]$).
We will consider this solution only in the range $H_0>-\left[\gamma(t_\text{i}-t)-\delta(t_\text{i}-t)^2\right]$, namely for expanding universe, assuming in general $|\gamma|\ll H_0^2\,,|\delta|\ll H_0^3$.

The scale factor is given by
\begin{equation}
a(t)=\frac{a_0}{6}t\left[6H_0-2\delta t^2-3t(\gamma-2 t_\text{i} \delta)+6t_\text{i} (\gamma-\delta t_\text{i} )\right]\,,\label{a2}
\end{equation}
where $a_0>0$ is a constant.
When $t$ is close to $t_\text{i}$ we obtain a quasi de Sitter solution for the inflation, but when $t$ increases the Hubble parameter decreases and we leave the primordial acceleration. 

If $\lambda\ll\rho$, Equation~(\ref{F}) with $\mathcal E=0$ reads as (\ref{F2})
and leads to
\begin{equation}
\rho\simeq\sqrt{\frac{6\lambda}{\kappa^2}}\left[H_0+(t_\text{i} -t)\left(\gamma-\delta(t_\text{i}-t)\right)\right]\,.\label{fluid2}
\end{equation}
This quantity is positive when $H$ is positive. In order to obtain this result, one simple choice may be to consider the following EoS for viscous fluid,
\begin{equation}
p=-\rho-\sqrt{\frac{6\lambda}{\kappa^2}}\frac{\dot H}{3 H}\,.
\end{equation}
However, this is not a good choice since in this way the conservation law (\ref{cons2}) results to be always satisfied for every solution of (\ref{F2}). An other possibility can be reconstructed as
\begin{equation}
p=-\rho-\left(\frac{6\lambda}{\kappa^2}\right)\frac{(2\delta f(\rho)-\gamma)}{3 \rho}\,,
\quad f(\rho)=
\frac{\gamma\pm\sqrt{\gamma^2-4\sqrt{\kappa^2/(6\lambda)}\rho\,\delta+4H_0\delta}}{2\delta}\,,\label{v2}
\end{equation}
which corresponds to (\ref{EoS}) with $\omega(\rho)=-1-(6\lambda/\kappa^2)(2\delta f (\rho)-\gamma)/(3\rho^2)$ and viscosity equal to zero (inhomogeneous non viscous fluid).

The slow roll parameters (\ref{epsilon})--(\ref{eta}) read
\begin{equation}
\epsilon=\frac{\gamma-2\delta(t_\text{i}-t)}{\left[H_0+(t_\text{i}-t)(\gamma-\delta(t_\text{i}-t))\right]^2}\,,
\quad\eta=-\frac{\delta}{\left[\gamma-2\delta(t_\text{i}-t))(H_0+(t_\text{i}-t)(\gamma-\delta(t_\text{i}-t))\right]}\,,
\end{equation}
and if
\begin{equation}
\gamma\ll H_0^2\,,\quad\delta\ll\gamma H_0\label{rel2}\,,
\end{equation}
they are very small when $t$ is close to $t_\text{i}$,
\begin{equation}
\epsilon\simeq\frac{\gamma}{H_0^2}\,,\quad |\eta|\simeq\frac{\delta}{\gamma H_0}\,.
\end{equation}
The $\epsilon$ slow roll parameter is on the order of the unit and inflation ends when
\begin{equation}
t_\text{f}\simeq t_\text{i}+\frac{H_0}{2\gamma}\,.
\end{equation}
We note that $t_\text{i}\ll t_\text{f}$ thanks to (\ref{rel2}). 

The $e$-folds number of inflation (\ref{N}) reads
\begin{equation}
N\simeq H_0 (t_\text{f}-t_\text{i})=\frac{H_0^2}{2\gamma}\,,
\end{equation}
where we have considered $t_\text{i}\ll t_\text{f}$. 
Thus, to obtain $76<N$ we must require $152\gamma<H_0^2$. Note that during the inflation
\begin{equation}
\epsilon\simeq\frac{1}{2N}\,.
\end{equation}
The amplitude of power spectrum (\ref{spectrum}) is given by
\begin{equation}
\Delta^2_\mathcal R=\frac{\kappa^2 H_0^4}{8\pi^2\gamma}\,.
\end{equation}
Finally, the spectral indexes (\ref{index}) are derived as
\begin{equation}
n_s=1-\frac{6\gamma}{H_0^2}+\frac{2\delta}{\gamma H_0}\sim1-\frac{2}{N}\,,\quad r=\frac{16\gamma}{ H_0^2}\simeq\frac{8}{N}\,.
\end{equation}
These quantities satisfy the Plank data, but can be also near to the last BICEP2 results.

We note that when $\rho\ll\lambda $, the fluid conservation law with the EoS (\ref{v2}), by choosing the sign plus to have the positivity of $f(H)$ and therefore the negativity of the second term in the pressure,  leads to $\rho\propto\sqrt{\log a/a_\infty}$, where $a_\infty$ can be set as the value of the scale factor when $t\rightarrow\infty$, and the fluid contribute disappears from cosmological scenario in expanding universe.

\subsection{Quintessence inflation\label{quintinfl}}

In this last subsection, we want to consider the quintessence inflation from viscous fluid with brane correction: the Hubble parameter decreases with the energy density of the fluid, which passes from the region $\lambda\ll\rho$ with Equation (\ref{F2}) where acceleration takes place, to the region $\rho\ll\lambda$ with the Friedmann Equation (\ref{Fnormal}), where the fluid induces deceleration and inflation ends. 

Let us assume the following form for the EoS (\ref{EoS}) of the homogeneous viscous fluid,
\begin{equation}
p=\omega_\text{eff}\rho-3H f(H)\zeta_0\,,\quad f(H)=\exp\left[-\frac{\kappa^2\lambda}{H^2}\right]\,,\quad\zeta_0=\sqrt{\frac{\lambda}{6\kappa^2}}(1+2\omega_\text{eff}-\omega)\,,\label{EoS3}
\end{equation} 
with 
\begin{equation}
-1<\omega<-\frac{1}{3}\,,\quad\frac{(\omega-1)}{2}<\omega_\text{eff}\,,
\end{equation}
where the second condition guarantees the positivity of the viscosity and $f(H)$ cut off the viscosity at low energy, when $\rho\ll\lambda$.
In the high energy limit $\lambda\ll\rho$, when $\kappa^2\lambda\ll H^2$, Equation (\ref{cons2}) with Equation (\ref{F2}) lead to
\begin{equation}
H(t)\simeq\frac{2}{3(1+\omega)t}\,,\quad
\rho\simeq\sqrt{\frac{6\lambda}{\kappa^2}}\left[\frac{2}{3(1+\omega)t}\right]\,,\label{H3}
\end{equation}
showing the initial singularity at $t=0$ which may be identified with the Big Bang. Due to this singularity, the beginning of the inflation is choosen at the Planck time, $t_\text{i}=t_{Pl}\sim 10^{-44}\text{sec.}$, when the acceleration starts.
Thus, the slow roll parameters (\ref{epsilon})--(\ref{eta}) are given by
\begin{equation}
\epsilon=\eta=\frac{3(1+\omega)}{2}\,,\label{slow3}
\end{equation}
and inflation is viable if $\omega$ is very close to minus one. The fact that the slow roll parameters are constant is not surprising, since quintessence universe gives an eternal acceleration. However, the energy density of the fluid decreases and the solution of Equation (\ref{F}) changes. In the specific, in the low energy limit $\rho\ll\lambda$, when $H^2\ll\kappa^2\lambda$ and Equation (\ref{F}) returns to be (\ref{Fnormal}), the solution reads
\begin{equation}
H(t)\simeq\frac{2}{3(1+\omega_\text{eff}) t}\,,\quad\rho(t)\simeq\rho_0 a^{-3(1+\omega_\text{eff})}\,.
\end{equation}
For example, by choosing $\omega_\text{eff}=1/3$, we may recover the radiation dominatated universe with $H=1/(2 t)$. It means that, due to the mechanism induced by the brane correction, all the energy of the fluid is transferred to the ultrarelativistic matter/radiation of Friedmann universe at the end of inflation, without reheating process. 

The inflation ends and the model passes to the low energy limit when $H^2\simeq\kappa^2\lambda/6$, namely at
\begin{equation}
t_\text{f}\simeq\frac{2}{3(1+\omega)}\sqrt{\frac{6}{\kappa^2\lambda}}\,,
\label{t3}
\end{equation}
which is large for $\omega$ close to minus one.

Since solution (\ref{H3}) is not a de Sitter, the expression for the $e$-folds number (\ref{N}) must be modified as
\begin{equation}
N\equiv\log\left(\frac{\dot a_f}{\dot a_i}\right)=\frac{2-3(1+\omega)}{3(1+\omega)}\ln\left(\frac{t_\text{f}}{t_\text{i}}\right)\,,
\end{equation}
and corresponds to (\ref{N}) when $\omega$ is close to minus one. 
In order to 
obtain $76<N$, one must require $\exp[114(1+\omega)]\ll(t_\text{f}/t_\text{i})$, where the duration of the inflation depends on the 3-brane tension $\lambda$ as in (\ref{t3}). 

The amplitude of power spectrum (\ref{spectrum}) for quintessence inflation is 
\begin{equation}
\Delta_\mathcal{R}^2=\frac{\kappa^2}{27\pi^2(1+\omega)^3t^2}\,,
\end{equation} 
and the spectral indexes (\ref{index}) read
\begin{equation}
n_s=(1-6(1+\omega))\,,\quad r=24(1+\omega)\,.
\end{equation}
In general, the Plank satellite results can be satisfied from the model, but in principle also the last BICEP2 results may be found: in the specific, if  $\omega\simeq-0.992$, we have $r\simeq0.20$, and the slow roll limit of inflation is well realized since $\epsilon, \eta\simeq 0.012$
are very small and inflation is viable.

\section{Late-time fluid evolution}

Now it could be interesting to spend several words about the late time evolution of the fluid inflation models. We already have seen in the preceeding examples the behaviour of the considered fluids after inflation, namely in the low energy limit of Equation (\ref{F}), when we recover the standard Friedmann equation (\ref{Fnormal}) and in general the fluids induce deceleration. 
We have stressed that in order to find the Friedmann universe some rehating process coming from oscillations of the fluid energy density and therefore of the curvature is necessary~\cite{mioultimo}; otherwise, the energy density of the fluid must be converted in radiation, like in the example considered in \S~\ref{quintinfl}. 

One additional interesting task could be the unification of the early- and late-time cosmological acceleration of the universe today in an unique fluid model. In standard cosmology, since the Friedmann equations keep the same form at every time, it is possible only if the fluid EoS changes the behaviour at different energies (Hubble parameter) but here, thanks to the brane correction, the Hubble parameter itself  changes its behaviour respect to the fluid energy density at different scales. 

We note that in the high energy limit (\ref{F2}), to produce the accelerated solution
\begin{equation}
H(t)=\frac{2}{3(1+\omega)t}\,,\quad-1<\omega<-1/3\,,\quad H(t)=\frac{2}{3(1+\omega)(t_0-t)}\,,
\quad\omega<-1\,,\label{HH}
\end{equation}
where we have introduced the singularity at the future time $t<t_0$ for the phantom case $\omega<-1$ in order to mantain the positivity of the Hubble parameter~\cite{Caldwell}, it is necessary a perfect fluid with the EoS
\begin{equation}
p=\omega_\text{eff}\rho\,,\quad\omega_\text{eff}=\frac{(\omega-1)}{2}\,.
\end{equation}
It means that in the presence of the brane correction a perfect fluid produces acceleration if the related EoS paramter $\omega_\text{eff}$ is 
\begin{equation}
\omega_{\text{eff}}<-\frac{2}{3}\,,
\end{equation}
and never brings deceleration at low energy with (\ref{Fnormal}), when $\omega$ in (\ref{HH}) shifts to $\omega_\text{eff}<-1/3$ and we still have acceleration. This is the reason for which in \S~\ref{quintinfl} we have introduced a viscosity into the fluid, changing the fluid EoS at low energy and converting its energy density in standard radition.
However, other possibilities are allowed. For example, we may transform the fluid for the early-time acceleration in a dark fluid for the present dark energy epoch with $\omega=-1$. 
If we use a more general form of the viscous EoS with the terms depending on Hubble parameter and energy density mixed together, we may write,
\begin{equation}
p=-\rho+\left(\frac{\rho}{\lambda}\right)\frac{\rho^2\kappa^2}{6H^2}(1+\omega)
\,,\quad-1<\omega<-\frac{1}{3}\,,
\end{equation}
such that when $\lambda\ll\rho$, if we use Equation (\ref{F2}) we find $p\simeq\omega\rho$ and we get the quintessence inflation, but when $\rho\ll\lambda$, the second term in the expression above is much smaller than $\rho$ and the EoS of the fluid reads $p\simeq-\rho$, with constant $\rho\simeq\rho_0$: by bounding such a constant to the observed value of the Cosmological constant $\Lambda\simeq 10^{-66}\text{eV}^{2}$ (with $\rho_0=\Lambda/\kappa^2$), we may obtain the late-time acceleration of the universe today.

\section{Conclusions}

In this work, we have analyzed some features of models of inhomogeneous viscous fluids for inflation by taking into account the brane correction comining from five dimensional theory of gravity in Friedmann-Robertson-Walker space-time. Braneworld inflation has well studied in literature in the canonical scalar field representation, where several differences with respect to the classical Einstein's case have been stressed: the slow roll parameters result to be very small also when the potential of the field is not so ``flat'' , the dependence of the $e$-folds number  on the $\epsilon$ slow roll parameter as $N\sim 1/\sqrt{\epsilon}$ is slightly modified from the contribution of the potential rescaled by the 3-brane tension and so on.

Inhomogeneous viscous fluid cosmology offers the possibility to reproduce a huge list of different scenarios by playing with the Equation of State of the fluids. In our case, we investigated three different kinds of inflation, namely inflation with exponential Hubble parameter in \S~\ref{expH}, quasi de Sitter inflation in \S~\ref{QuasidS}, and quintessence inflation in \S~\ref{quintinfl}. We have shown how the fluid models can induce acceleration at high energy when the brane correction to Einstein's gravity becomes important, and give deceleration at low energies. 
In particular, in the last example of quintessence inflation, we have seen how in principle it is possible to convert the energy density of the fluid in ultrarelativistic matter/radiation after the end of inflation without reheating. All the proposed scenarios describe a viable inflation. A general result that we have found throught our analysis is that the slow roll parameters $\epsilon\,,\eta$ and therefore the spectral indexes of the fluid models are bigger than the ones of scalar field theories of inflation, confirming the results of Ref.~\cite{mioultimo} for standard Einstein's gravity. Here, the $e$-folds number is proportinal to $1/\epsilon$ instead of $1/\sqrt{\epsilon}$ and the amount of inflation can be enough large even if $\epsilon$ is not vanishing. As a consequence, the fluid models for inflation may reproduce the results of last BICEP2 esperiment where the tensor-to-scalar ratio is not so close to zero. 

In the end, we also stressed that fluid models may be investigated in the attempt to find an unified scenario for early- and late-time cosmological acceleration and in the last section we have given a qualitative example.

Other important works about generalizations of braneworld cosmology can be found in Refs.~\cite{superbrane1, superbrane2}.
Some works on fluid cosmology and the dark energy issue have been also developed in Ref.~\cite{Ciappi}, in Refs.~\cite{Carro}--\cite{undici}, in Ref.~\cite{LittleRip} for viscous fluids in Little Rip cosmology and in Ref.~\cite{pp} for fluid perturbations in FRW space-time.

\section*{Acknowledgments}

We would like to thank Professor Sami for valuable comments and interesting discussions.

\end{document}